\documentclass[12pt,fleqn]{article}
\usepackage{exscale}
\usepackage{epsfig}
\begin{document}

\title{Exact solution of the multi-component Falicov-Kimball model
        in infinite dimensions}

\author{V. Zlati\'c$^{1,2}$, J. K. Freericks$^{1}$, 
        R. Lemanski$^{1,3}$, and G. Czycholl$^{4}$     \\[2ex]
\em $^1$ Department of Physics, Georgetown University, \\
\em          Washington, DC 20057, USA,       \\  
\em $^2$ Institute of Physics, P.O.B 304,     \\
\em         10 001 Zagreb, Croatia,           \\
\em $^3$Institute of Low Temperatures and Structure Research, \\
\em     Polish Academy of Sciences, Wroc\l aw, Poland          \\
\em $^4$ Intitut f\"ur Theoretische Physik, Universit\"at Bremen, \\
\em       D-28334 Bremen, Germany}

\date{\today}
\maketitle

\begin{abstract}
The exact solution for the thermodynamic and dynamic properties of the 
infinite-dimensional multi-component Falicov-Kimball model for 
arbitrary concentration of d- and f-electrons is presented. 
The emphasis is on a descriptive derivation of important physical 
quantities by the equation of motion technique.
We provide a thorough discussion of the f-electron Green function and
of the susceptibility to spontaneous hybridization. The solutions are
used to illustrate different physical systems ranging from the
high-temperature phase of the YbInCu$_4$ family of materials to
an examination of classical intermediate valence systems that can develop
a spontaneous hybridization at $T=0$.

\end{abstract}

\section*{Introduction}
The anomalous features observed in the YbInCu$_4$ family 
of intermetallic compounds (Sarrao et al 1999) seem to be driven 
by the short-range Coulomb interaction between highly-degenerate 
Yb f-holes and the conduction states (Freericks and Zlati\'c 1998). 
The same interaction seems to be responsible for the 
optical anomalies in SmB$_6$ and related compounds (such as
correlated ferroelectrics) 
(Wachter and Travaglini 1985, Guntherodt et al 1982, Portengen et al 1996).
We study the effect of this interaction using the multi-component  
Falicov-Kimball model (Falicov and Kimball 1969) in infinite dimensions, 
where all the thermodynamic properties can be calculated exactly. 
The model consists of $(2s+1)$--fold degenerate mobile d-electrons 
and static $(2S+1)$--fold degenerate f-electrons, 
which interact via an onsite Coulomb interaction $U$. 
The model was originally introduced to describe metal-insulator transitions
in materials that do not change the character of their electronic states, but 
do change their thermodynamic occupations as functions of the 
external parameters.
The exact results for the static and dynamic correlation functions 
of the spin-one-half model explain the collapse of the low-temperature 
metallic phase of YbInCu$_4$-like systems, and account in a qualitative 
way for most of their features in the paramagnetic, semiconducting, 
high-temperature phase. 
The exact solution of the spinless model shows that the statistical 
fluctuations give rise to a logarithmic divergence (in $T$)
of the spontaneous hybridization correlation 
function at zero temperature, so that any amount of quantum mixing 
could lead to a phase transition at finite temperatures. This might be 
relevant for SmB$_6$ and for correlated ferroelectrics (Portengen et al 1996).

In what follows, we describe the model, explain the method of 
calculating the Green's functions for d- and f-electrons, and present 
results for some static and dynamic correlation functions 
of a spinless and spin-degenerate case. Detailed comparison with 
the experimental data will be given elsewhere. 
 
\section*{Formalism for the d-electron Green's function}

The multi-component Falicov-Kimball model (Falicov and Kimball 1969) 
describes the dynamics of two types of electrons:  
the conduction electrons (created or destroyed at site $i$ by 
$d_{i\sigma}^{\dagger}$ or $d_{i\sigma}$) 
and localized electrons (created or destroyed at site $i$ by 
$f_{i\eta}^{\dagger}$ or $f_{i\eta}$). 
The $(2s+1)$--fold degenerate d-states and the $(2S+1)$--fold degenerate 
f-states are labeled by $\sigma$ and $\eta$, respectively.
The  multi--component model is used to describe the electrons with 
spin and/or orbital degrees of freedom, and $2s+1$ and $2S+1$ can assume 
different values. The non-interacting conduction electrons can hop 
between nearest-neighbor sites on the D-dimensional lattice, 
with a hopping matrix  $-t_{ij}=-t^*/2\sqrt{D}$; 
we choose a scaling of the hopping matrix that yields a nontrivial limit
in infinite-dimensions (Metzner and Vollhardt 1989).  
The f-electrons have a site energy $E_f$, 
and a chemical potential $\mu$ is employed to conserve the 
total number of electrons $n_d+n_f=n_{tot}$. 
The d- and f-number operators at each site are 
$n_d=\sum_{\sigma}n_{d\sigma}$ and $n_f=\sum_{\eta}n_{f\eta}$. 
We assume an infinite Coulomb repulsion between f-electrons 
with different label $\eta$, and restrict the f-occupancy at a given 
site to $n_f\leq 1$, regardless of the degeneracy of the f-state.
The Coulomb interaction $U$ between the d- and f-electrons
that occupy the same lattice site is finite, and the 
Falicov-Kimball (FK) Hamiltonian for the lattice 
is defined as (Brandt and Mielsch 1989, Freericks and Zlati\'c 1998),
\begin{equation}
H_{FK}=
\sum_{ij,\sigma }(-t_{ij}-\mu \delta _{ij})d_{i\sigma }^{\dagger}d_{j\sigma }
+
\sum_{i,\eta }(E_f-\mu )f_{i\eta }^{\dagger }f_{i\eta }
+
U\sum_{i,\sigma \eta}
d_{i\sigma }^{\dagger }d_{i\sigma}
f_{i\eta}^{\dagger }f_{i\eta}. 
                                                  \label{H_FK}
\end{equation}
The FK lattice model (\ref{H_FK}) can be solved in infinite 
dimensions using the methods of Brandt and Mielsch (Brandt and Mielsch 1989).
We consider two kinds of lattices: (i)
the hypercubic lattice with a Gaussian noninteracting density 
of states $\rho(\epsilon)
=
\exp [-\epsilon^2/t^{*2}]/(\sqrt{\pi}t^*)$; and (ii) the infinite-coordination
Bethe lattice with a semicircular noninteracting density of states
$\rho(\epsilon)=\sqrt{4t^{*2}-\epsilon^2}/(2\pi t^{*2})$;
and we take $t^*$ as the unit of energy ($t^*=1$). 
The method of calculation is formulated for arbitrary values of s and 
S labels, but the results are presented for the spinless model 
($2s+1=1$ and $2S+1=1$), and for the spin-one-half model 
($2s+1=2$ and $2S+1=2$). We consider only the homogeneous phase, 
where all quantities are translationally invariant. 
\subsubsection*{Mapping onto the Falicov-Kimball atom}
Infinite-coordination lattices can be solved by a mean-field-like  
procedure, because the self energy of the conduction electrons is 
local (Metzner and Vollhardt 1989). That is, the Dyson's equation for the local 
d-electron Green's function $G_{loc}^{\sigma}(z)$ on the lattice reads
\begin{equation}
G_{loc}^{\sigma}(z) 
=\int \frac{\rho(\epsilon)}
{z+\mu-\Sigma^{\sigma}(z)-\epsilon}d\epsilon,
                                            \label{eq: gloc}
\end{equation}
where $z$ is a complex variable, and $\Sigma^{\sigma}$ is the 
momentum-independent self energy.
Hence, as noted by Brandt and Mielsch (Brandt and Mielsch 1989), the lattice 
self energy coincides with the self energy of an atomic d-state 
coupled to an f-state by the same Coulomb interaction as on the 
lattice,  and perturbed by an external time-dependent field, 
$\lambda^\sigma(\tau,\tau^{\prime})$. 
For an appropriate choice of the $\lambda$-field, 
the functional dependence of $\Sigma^{\sigma}$ on $G^{\sigma}(z)$ 
and $F^{\sigma}(z)$, the atomic propagators for d- and f-states, 
is exactly the same as in the lattice case. 
The lattice problem is thus reduced to finding the atomic 
self-energy functional for d-electrons, and then setting 
$G_{loc}^{\sigma}(z) = G^{\sigma}(z)$ and 
$F_{loc}^{\sigma}(z) = F^{\sigma}(z)$ at each lattice site. 

The FK atom can be solved by using the interaction representation, 
such that the time dependence of operators is defined by the atomic 
Hamiltonian, 
\begin{equation}
                                         \label{H_atom}
H_{\rm at}
=
 -
 \mu \sum_{\sigma }d_{\sigma }^{\dagger }d_{\sigma }
 +
 (E_f-\mu )\sum_{\eta }f_{\eta }^{\dagger }f_{\eta}
 +
 U\sum_{\sigma \eta}
 d_{\sigma}^{\dagger }d_{\sigma}
 f_{\eta}^{\dagger }f_{\eta}, 
  \end{equation}
and the time-dependence of the state vectors is governed by the 
evolution operator,
\begin{equation}
                                               \label{S-matrix}
S(\lambda)
=
T_{\tau}
e^{-\int_0^{\beta}d\tau\int_0^{\beta} d\tau^{\prime}
     \sum_\sigma \lambda^\sigma(\tau,\tau^{\prime})
       d_{\sigma }^{\dagger}(\tau)d_{\sigma}(\tau^{\prime})} .
\end{equation} 
The external field $\lambda^\sigma(\tau,\tau^{\prime})$ is 
assumed to be time-translation invariant in (imaginary) time,  
\begin{equation}
                                               \label{lambda-Fourier}
 \lambda^\sigma(\tau,\tau^{\prime})
 = 
 T \sum_n
 \lambda_n^\sigma e^{-i\omega_n (\tau -\tau^{\prime})} ,
\end{equation}
and hence can be expanded in a Fourier series in the Fermionic Matsubara
frequencies $\omega_n=\pi(2n+1)T$, where we set $k_B=1$.  
In the absence of an external magnetic field, 
the $\lambda$--field is the same for all the $\sigma$--components.
The unperturbed atomic Hamiltonian (\ref{H_atom}) conserves the number 
of f- and d-electrons, while the time dependent $\lambda$-field 
gives rise to fluctuations in the d-occupancy. In the equivalent 
lattice problem, the local d-fluctuations are due to the 
d-electron hopping. 

The thermodynamics of the FK atom follows from the partition function, 
 \begin{equation}
                                                 \label{Z}
 {\cal Z}_{at}(\lambda)
 ={\rm Tr}_{df} \left[
 T_{\tau}
 e^{-\beta H_{at}} S(\lambda)
 \right], 
 \end{equation}
where the statistical sum runs over all possible atomic 
configurations, which is determined by function 
$\lambda^\sigma(\tau,\tau^{\prime})$ for $\tau,\tau^\prime \in (0,\beta)$. 
The specific feature of the atomic FK model is that the 
number of f-electrons is a constant of motion, 
and is either zero or one, while the d-electron number 
can  assume any value between zero and $2s+1$. 
Furthermore, the evolution operator  $S(\lambda)$ never transfers the state 
vectors out of the invariant ($n_f=0$ or $n_f=1$) Hilbert subspaces, so  
the matrix elements can be evaluated within each invariant subspace by 
replacing the operator $\sum_{\eta}f_{\eta}^{\dagger }f_{\eta}$ by its 
eigenvalue 0 or 1. The trace in ({\ref{Z}) can thus be performed separately 
for the $n_f=0$ and $n_f=1$ subspaces.  
Within the $n_f=0$ subspace, the operator dynamics is governed by a 
simplified non-interacting Hamiltonian, 
\begin{equation}
                                      \label{ham_nonint}
 H_{0}
=
 -\mu \sum_{\sigma }d_{\sigma }^{\dagger }d_{\sigma }, 
\end{equation}
and we have $d_{\sigma}^\dag(\tau) = d_{\sigma}^\dag \exp(-\mu\tau)$
and  $d_{\sigma}(\tau)=d_{\sigma} \exp(\mu\tau)$, 
where $d_{\sigma}^\dag=d_{\sigma}^\dag(0)$ and $d_{\sigma}=d_{\sigma}(0)$ 
are the time-independent Schroedinger operators.
The operator dynamics in the $n_f=1$ subspace is governed by the 
same Hamiltonian as in the $n_f=0$ subspace but with $\mu$ 
replaced by $\mu-U$. The trace over f-states in (\ref{Z}) can be performed, 
and we find, 
\begin{equation}
                                                 \label{Z_Z0}
 {\cal Z}_{at}(\lambda)={\cal Z}_0(\lambda,\mu) 
 +
 (2S+1)e^{-\beta (E_f-\mu)}{\cal Z}_0(\lambda,\mu-U),
\end{equation}
where
\begin{equation}
                                             \label{Z0_product}
{\cal Z}_0(\lambda)
 =
\prod_\sigma {\cal Z}_0^\sigma(\lambda^\sigma),
 \end{equation}
and 
 \begin{equation}
                                             \label{Z0-spin}
{\cal Z}_0^\sigma(\lambda^\sigma)
 ={\rm Tr}_d \left[
 T_{\tau}
 e^{-\beta H_0^\sigma} S(\lambda^\sigma)
 \right] .
 \end{equation}
The factorization (\ref{Z0_product}) holds 
because the time evolution due to $H_0$ is such that the 
operators with different $\sigma$-labels commute regardless 
of their time arguments, and the S-matrix (\ref{S-matrix}) 
does not change the $\sigma$-label of a given state vector.  
Thus, the Hilbert space can be decomposed into invariant 
$\sigma$--subspaces and the trace in (\ref{Z0-spin}) is over 
the non-degenerate d$_\sigma$--states. 
In each of these subspaces, the operator dynamics is defined by 
$H_{0}^\sigma$, and the partition function ${\cal Z}_0^\sigma(\lambda^\sigma)$ 
describes the statistics of a d$_\sigma$--electron subject 
to an arbitrary time-dependent $\lambda^\sigma$-field. 

In the presence of the magnetic field, the Hamiltonians (\ref{H_FK}) 
and (\ref{H_atom}) have to be supplemented with Zeeman terms, 
\begin{equation}
                                         \label{H_Zeeman}
H_{\rm Z}
=
 g_d\mu_B \sum_{\sigma} \sigma d_{\sigma}^{\dagger }d_{\sigma}
+
 g_f\mu_B \sum_{\eta} \eta f_{\eta}^{\dagger }f_{\eta}, 
\end{equation}
where $g_d$ and $g_f$ are the g-factors of d- and f-electrons, respectively. 
The solution of the model for $H\neq 0$ is a straightforward 
generalization of the $H=0$ case, which is presented here (the main difference
is that the effective chemical potentials now have a spin index dependence). 
\subsubsection*{Generalized partition function for the Falicov-Kimball atom}
The self-energy functional for the FK atom is calculated by using 
the equations of motion (EOM) for the Green's function 
obtained by functional differentiation of the generalized partition 
function ${\cal Z}_{at}(\lambda)$ (Kadanoff and Baym 1962). 
Equations (\ref{Z_Z0}) to (\ref{Z0-spin}) show that we can find  
${\cal Z}_{at}(\lambda)$ by solving a statistical problem for a single 
non-degenerate d$_\sigma$-state coupled to the periodic $\lambda^\sigma$--field.
Consider the contribution to ${\cal Z}_0^\sigma(\lambda^\sigma)$  
due to the shift of the $\lambda$-field from an initial configuration 
$\lambda^\sigma(\tau,\tau^{\prime})$ to the final configuration  
$\lambda^\sigma(\tau,\tau^{\prime})+\delta\lambda^\sigma(\tau,\tau^{\prime})$,
\begin{equation}
                                               \label{Z_0-variation}
\delta\,{\cal Z}_0^\sigma
=
{\rm Tr}_{d} 
\langle 
T_{\tau}
e^{-\beta H_{0}^\sigma} 
\delta S(\lambda^\sigma) 
\rangle.
\end{equation}
The $\delta S(\lambda^\sigma)$ is obtained from 
the usual rules of the calculus of variations,
\begin{eqnarray}
                                               \label{delta-S}
\delta \, S(\lambda^\sigma)
&=&
\delta\, \exp
   \left\{ 
      -
      \int_0^{\beta}d\tau\int_0^{\beta} d\tau^{\prime}
      \lambda^\sigma(\tau,\tau^{\prime})
       d_{\sigma }^{\dagger}(\tau)d_{\sigma}(\tau^{\prime})
   \right\}
                                               \nonumber\\
&=&
- S(\lambda) 
\int_0^{\beta}d\tau\int_0^{\beta} d\tau^{\prime}
      \delta\lambda^\sigma(\tau,\tau^{\prime})
       d_{\sigma }^{\dagger}(\tau)d_{\sigma}(\tau^{\prime}) ,
\end{eqnarray}
and the time-ordering is taken into account when this result is
substituted into (\ref{Z_0-variation}).
Performing the substitution  gives 
\begin{equation}    
                                               \label{delta-Z_0}
\delta\,\ln {\cal Z}_0^\sigma
=
\int_0^{\beta}d\tau\int_0^{\beta} d\tau^{\prime}
      \delta\lambda^\sigma(\tau,\tau^{\prime})
       G^{\sigma}_{0}(\tau^\prime,\tau), 
\end{equation}
where 
\begin{equation}
                                                 \label{G_0}
G^{\sigma}_{0}(\tau,\tau^\prime)
= -\frac{1}{{\cal Z}_0^\sigma}
{\rm Tr}_{d} \left\{
T_{\tau}
e^{ -\beta H_0}
d_{\sigma}(\tau) d_{\sigma }^{\dagger}(\tau^\prime)
S(\lambda) \right\}, 
\end{equation}
is the d-electron Green's function restricted to configurations with no
f-electrons.
The function multiplying the variation $\delta\lambda^\sigma(\tau^{\prime},\tau)$ 
is, by definition, the functional derivative of the partition function, 
\begin{equation}
                                            \label{G_0-definition}
G_{0}^{\sigma}(\tau,\tau^\prime) 
=
 -\frac
{\delta \ln{\cal Z}_0^\sigma}{\delta \lambda^{\sigma}(\tau^\prime,\tau)}.
\end{equation}
Expressing $G_{0}^{\sigma}(\tau,\tau^\prime)$ in Eq.~(\ref{delta-Z_0}) 
in terms of its Fourier components, 
\begin{equation}
                                               \label{G_0_Fourier}
G_{0}^{\sigma}(\tau,\tau^\prime) 
=
T \sum_n  
G_{0n}^{\sigma}
e^{-i\omega_n (\tau -\tau^{\prime})}. 
\end{equation}
and using (\ref{lambda-Fourier}) for 
$ \delta\lambda^\sigma(\tau,\tau^{\prime})$ 
leads to the functional relation, 
\begin{equation}    
                                               \label{delta-Z_0-Fourier}
\delta\,\ln {\cal Z}_0^\sigma
=
\sum_n
G_{0n}^{\sigma}\delta\lambda_{n}^{\sigma}
\end{equation}
where $G_{0n}^{\sigma}$ is now defined as a simple partial derivative,  
\begin{equation}
                                               \label{G_0-Z_0-Fourier}
G_{0n}^{\sigma} 
=
 -\frac{\partial \ln {\cal Z}_0^\sigma}{\partial \lambda_n^\sigma}. 
\end{equation}
An explicit expression for $G_{0n}^{\sigma}(\lambda_{n}^{\sigma})$ 
would allow us to obtain ${\cal Z}_0^\sigma$ by  solving the 
differential equation (\ref{G_0-Z_0-Fourier}). 
Functions $G_{0n}^{\sigma}$ and $\lambda_n^\sigma$, or 
$G_{0}^{\sigma}(\tau,\tau^\prime)$ and $\lambda^\sigma(\tau^{\prime},\tau)$,  
can be considered as matrix elements of integral operators 
$G_{0}^{\sigma}$ and $\lambda^\sigma$, and  Eqs.~(\ref{delta-Z_0}) 
and (\ref{G_0-Z_0-Fourier}) can be written in the operator form as, 
\begin{equation}    
                                               \label{delta-operators}
\delta\,\ln {\cal Z}_0^\sigma
= {\rm Tr}~ \left\{ G_0^\sigma \delta\lambda^\sigma\right\},
\end{equation}
where the trace denotes an integration over time if we are using 
non-diagonal matrices in the $\tau$--representation, or a Matsubara summation, 
if we are using diagonal matrices in  frequency space. 

Our next step is to find an explicit expression for $G_{0n}^{\sigma}$,
and solve (\ref{G_0-Z_0-Fourier}) to find ${\cal Z}_0^\sigma$. 
The Green's function in Eqs.~(\ref{G_0}) and (\ref{G_0-Z_0-Fourier}) 
are obtained by the EOM.  Consider first the case $\tau >\tau^\prime$.
To compute $\partial /\partial\tau G^{\sigma}_{0}(\tau,\tau^\prime)$, we must
first compute the derivative of $[T_\tau d_\sigma(\tau)d_\sigma^{\dagger}
(\tau^\prime)S(\lambda^\sigma)]$ with respect to $\tau$.  
It is important to note that the differential operator does not 
commute with the time-ordering operator, so one must
proceed carefully.  Note that when we take a derivative with respect to
$S(\lambda^\sigma)$ it will bring down terms 
like $d_\sigma^{\dagger}(\tau)d_\sigma (\tau_2)$ or
$d_\sigma^{\dagger}(\tau_1)d_\sigma(\tau)$, and the latter terms will not 
contribute
when multiplied by $d_\sigma(\tau)$---that is, the time-ordering with respect to
$\tau_1$ is the only important variable to consider when taking the derivative.
So we write the full time ordered product in two pieces
\begin{equation}
T_\tau d_\sigma(\tau)d_\sigma^{\dagger}(\tau^\prime)S(\lambda^\sigma)
=
[T_\tau \bar S(\lambda^\sigma)]
d_\sigma(\tau)
[T_\tau d_\sigma^{\dagger}(\tau^\prime)\bar{\bar S}(\lambda^\sigma)],
\end{equation}

with
\begin{eqnarray}
\bar S(\lambda^\sigma)
&=&
\exp \left [ -\int_\tau^\beta d\tau_1 \int_0^\beta d\tau_2
\lambda^\sigma(\tau_1,\tau_2)d_\sigma^{\dagger}(\tau_1)d_\sigma(\tau_2)\right],\cr
\bar{\bar S}(\lambda^\sigma)
&=&
\exp \left [ -\int_0^\tau d\tau_1 \int_0^\beta d\tau_2
\lambda^\sigma(\tau_1,\tau_2)d_\sigma^{\dagger}(\tau_1)d_\sigma(\tau_2) \right ] .
\end{eqnarray}
Now the derivative can be computed directly to yield
\begin{equation}
                                                \label{KB_5.6-5}
\frac{\partial}{\partial \tau}
T_{\tau}
\left[
  d_{\sigma}(\tau)d_{\sigma}^{\dagger}(\tau^\prime)S(\lambda^\sigma)
\right]
= T_\tau
\left[ \{
 \mu  d_{\sigma}(\tau)
 -\int_0^{\beta} d\tau_2
      \lambda^\sigma(\tau,\tau_2) d_{\sigma}(\tau_2)
\} d_{\sigma}^{\dagger}(\tau^\prime)
S(\lambda^\sigma) \right],
\end{equation}
where we employed identities like $d_\sigma(\tau_2)d_\sigma^\dagger(\tau)
d_\sigma(\tau)=d_\sigma(\tau_2)$ for $\tau_2>\tau$ (which can be easily derived
from the fact that the time dependence of the operators involve only an
exponential factor and the anticommutator of the Fermionic operators is 1).
Since the form for the Green's function is different for $\tau<\tau^\prime$,
one must repeat the derivation there (with the same result).  Hence 
the Green's function satisfies the following EOM, 
\begin{equation}
                                           \label{EOM}
(-\frac{\partial }{\partial\tau} + \mu)
G^{\sigma}_{0}(\tau,\tau^\prime)
-\int_0^\beta d \tau^{\prime\prime}
\lambda^\sigma(\tau,\tau^{\prime\prime})
G^{\sigma}_{0}(\tau^{\prime\prime},\tau^\prime)
=  \delta(\tau-\tau^\prime), 
\end{equation}
where the $\delta$-function arises from the discontinuity 
in $G_0$ at $\tau=\tau^\prime$. This EOM can also be written as, 
\begin{equation}
                                        \label{EOM-integral}
\int_0^\beta d \tau^{\prime\prime}
[G^{\sigma}_{0}]^{-1}(\tau,\tau^{\prime\prime})
G^{\sigma}_{0}(\tau^{\prime\prime},\tau^\prime)
=  \delta(\tau-\tau^\prime), 
\end{equation}
or, in matrix representation, 
\begin{equation}
                                        \label{integral-operator}
[G^{^\sigma}_{0}]^{-1}G^{^\sigma}_{0}= \mathbf{1},
\end{equation}
where $[G^{\sigma}_{0}]^{-1}$ is the non-diagonal integral operator 
defined by its matrix elements,  
\begin{equation}
                                             \label{G_0^(-1)_tau}
[G^{\sigma}_{0}]^{-1}(\tau,\tau^{\prime})
=
(-\frac{\partial }{\partial\tau} + \mu)
\delta(\tau-\tau^\prime)
-
\lambda^\sigma(\tau,\tau^{\prime}),
\end{equation}
and the unit operator $\mathbf{1}$ has the matrix elements
$\delta(\tau-\tau^\prime)$. 
The  Fourier transform of Eq.~(\ref{G_0^(-1)_tau}) gives the 
matrix elements of $[G^{\sigma}_{0}]^{-1}$ as, 
\begin{equation}
                                        \label{G_0^(-1)_omega}
[G^{\sigma}_{0}]^{-1}(i\omega_n)
={i\omega_n+\mu - \lambda^{\sigma}_n}, 
\end{equation}
and of its inverse $[G^{\sigma}_{0}]$ as,  
\begin{equation}
                                        \label{G_0_omega}
G^{\sigma}_{0}(i\omega_n)
=
\frac{1}{i\omega_n+\mu - \lambda^{\sigma}_n}.
\end{equation}
The diagonal form of $[G^{\sigma}_{0}]^{-1}$ and $[G^{\sigma}_{0}]$ 
in Fourier space is the consequence of the 
time-translation invariance of $\lambda^\sigma(\tau,\tau^{\prime})$. 
Using Eqs.~(\ref{G_0-Z_0-Fourier}) and (\ref{G_0_omega}) we  obtain 
the differential equation (Kadanoff and Baym 1962), 
\begin{equation}
                                                \label{dif_0}
 \frac{1}{i\omega_n+\mu - \lambda^{\sigma}_n}=
 -\frac{\partial \ln {\cal Z}_0^\sigma}{\partial \lambda_n^\sigma}, 
 \end{equation}
which has to be solved together with the initial ($\lambda=0$) boundary condition, 
 \begin{equation}
 {\cal Z}_0^\sigma(0,\mu)
 =1 + e^{\beta\mu} . 
 \end{equation}
The partition function for the simplified atomic problem 
is thus obtained as (Brandt and Mielsch 1989), 
\begin{equation}
                                        \label{solution_0}
 {\cal Z}_0^\sigma(\lambda^\sigma,\mu) 
 = 
2e^{\beta\mu/2}
  \prod_n \frac{i\omega_n+\mu - \lambda^{^\sigma}_n}{i\omega_n}.
\end{equation}
and the complete partition function for the FK atom can be written as, 
\begin{equation}
                                                 \label{Z_Z0-product}
 {\cal Z}_{at}(\lambda)
=
\prod_\sigma {\cal Z}_0^\sigma(\lambda^\sigma,\mu)
+ (2S+1)e^{-\beta (E_f-\mu)}
\prod_\sigma {\cal Z}_0^\sigma(\lambda^\sigma,\mu-U)
,
\end{equation}
\subsubsection*{Dynamics of the atomic d-state}
The renormalized d-electron propagator can be obtained, in complete analogy 
with Eqs.~(\ref{Z_0-variation})-(\ref{G_0-definition}), as a functional 
derivative of ${\cal Z}_{at}$ with respect to the external field, 
\begin{equation}
                                            \label{G-definition}
 G^{\sigma}(\tau,\tau^\prime) 
=
 -\frac{\delta \ln {\cal Z}_{at}}{\delta \lambda^{\sigma}(\tau^\prime,\tau)}.
\end{equation}
such that, 
\begin{equation}
                                               \label{G}
G^{\sigma}_{\rm at}(\tau,\tau^\prime)
= -\frac{1}{\cal Z}_{at}
{\rm Tr}_{df} \left<
T_{\tau}
e^{-\beta H_{\rm at}}
d_{\sigma}(\tau) d_{\sigma }^{\dagger}(\tau^{\prime})
S(\lambda^\sigma) \right>. 
\end{equation}
The difference with respect to the $n_f=0$ case is that the trace 
extends now over the d- and f-states, including the $\sigma$ and $\eta$ labels, 
and the statistical operator is the full atomic Hamiltonian $H_{at}$ 
rather than $H_0^\sigma$. 
On the imaginary frequency axis we still have, 
\begin{equation}
                                               \label{Z-G}
G_{n}^{\sigma} =
 -\frac{\partial \ln {\cal Z}_{at}}{\partial \lambda_n^{\sigma}},
\end{equation}
which gives, using Eqs.~(\ref{Z_Z0}) and (\ref{solution_0}), the result 
\begin{equation}
G_{n}^{\sigma}=
\frac{w_0}{[G^{\sigma}_{0n}]^{-1}}
+ 
\frac{w_1}{[G^{\sigma}_{0n}]^{-1}-U},
                                             \label{G-atomic}
\end{equation}
where $w_0 = {\cal Z}_0/{\cal Z}_{at}$ and $w_1=1-w_0$. The weight $w_1$ 
gives the f-occupation number (Brandt and Mielsch 1989). 

On the other hand, starting from the definition of the Green's function 
in (\ref{G}), and making the usual Feynman-Dyson expansion  with $U$ as 
the expansion parameter, we obtain the standard Feynman diagrams, 
in terms of the unperturbed propagators $G_{0n}^{\sigma}(\tau,\tau^\prime)$. 
The self-energy function of the FK atom $\Sigma_{n}^{\sigma}$ 
on the imaginary frequency axis is defined by the Dyson equation, 
\begin{equation}
\Sigma_{n}^{\sigma}
=
[G_{0n}^{\sigma}]^{-1} 
-
[G_{n}^{\sigma}]^{-1} . 
                                           \label{Dyson}
\end{equation}
Eliminating $G_{0n}^{\sigma}$, and hence the $\lambda$-field,  
from Eqs.~(\ref{G-atomic}) and (\ref{Dyson}) yields, 
\begin{equation}
                                             \label{sigma-CPA}
\Sigma_n^{\sigma} 
= 
\frac{w_1U}{1-(U-\Sigma_n^{\sigma})G_{n\sigma}}
\end{equation}
which can also be written in the form given by 
Brandt and Mielsch (Brandt and Mielsch 1989),
\begin{equation}
                                             \label{sigma-BM}
\Sigma_n^{\sigma} 
=
\frac{1}{2}\left(U-\frac{1}{G_{n\sigma}}
\pm \sqrt{\left(\frac{1}{G_{n\sigma}}-U\right)^2+4w_1\frac{U}{G_{n\sigma}}}
\right),
\end{equation}
where the sign of the square root is chosen to maintain the proper
analyticity properties of $\Sigma$. 

To clarify the physical meaning of the self-energy (\ref{sigma-CPA}) 
we now perform the calculations directly for the original lattice model 
(\ref{H_FK}) using the EOM. Since we also have to consider 
the Green's function of higher order in the creation and annihilation 
operators, it is convenient to introduce the compact Zubarev notation, 
where the Fourier transformed quantities are written as, 
\begin{equation}
\ll A , B \gg_n = \int_0^{\beta} d(\tau-\tau') 
e^{i \omega_n (\tau - \tau')} \langle T_{\tau} A(\tau) B(\tau') \rangle, 
\end{equation}
and $\langle \ldots \rangle$ denotes the thermal averaging over all the 
states on the lattice, 
\begin{equation}
                                                 \label{zubarev}
\langle \hat O \rangle
=
\frac{1}{ {\cal Z} }
{\rm Tr}_{df} 
e^{ -\beta H}
\hat O. 
\end{equation}
The Fourier transform of 
the lattice Green's function at site $i$ is now written 
as $G_{n\sigma}=\ll d_{i\sigma},d_{i\sigma}^{\dagger}\gg_n$, and the EOM reads, 
\begin{equation}
(\mu+i\omega_n)\ll d_{i\sigma},d_{i\sigma}^{\dagger} \gg_n 
= 1 
- \sum_j t_{ij} 
     \ll d_{j\sigma}, d_{i\sigma}^{\dagger} \gg_n 
+ U \sum_{\eta} 
   \ll d_{i\sigma} f_{i\eta}^{\dagger}f_{i\eta}, d_{i\sigma}^{\dagger}\gg_n .
                       \label{eomGerd1}
\end{equation}
Using the EOM for the higher order Green's function on the r.h.s. of 
Eq.~(\ref{eomGerd1}), and considering also the time derivative with 
respect to the second time-variable $\tau'$, we obtain
\begin{eqnarray}
(\mu + i \omega_n) \ll d_{i\sigma} f_{i\eta}^{\dagger} f_{i\eta} , 
d_{i\sigma}^{\dagger} \gg_n 
&=& 
\langle f_{i\eta}^{\dagger} f_{i\eta} \rangle
- \sum_j t_{ij} \ll d_{i\sigma} f_{i\eta}^{\dagger} f_{i\eta} , 
d_{j\sigma}^{\dagger} \gg_n \nonumber \\
&+& U \ll d_{i\sigma} f_{i\eta}^{\dagger} f_{i\eta} , 
 f_{i\eta}^{\dagger} f_{i\eta} d_{i\sigma}^{\dagger} \gg_n . 
\label{eomGerd2}
\end{eqnarray}
Here, we used the fact that a given site $i$ cannot be occupied by more 
than one f-electron, which means that the same spin index $\eta$ appears for the
f-electron operators at both times $\tau$ and $\tau'$. As the
f-occupation per site is conserved, we also have the relation
\begin{equation}
\ll d_{i\sigma} f_{i\eta}^{\dagger} f_{i\eta} , 
 f_{i\eta}^{\dagger} f_{i\eta} d_{i\sigma}^{\dagger} \gg_n =
\ll d_{i\sigma} f_{i\eta}^{\dagger} f_{i\eta} , 
d_{i\sigma}^{\dagger} \gg_n .
\label{eomGerd3} 
\end{equation}
Defining the local (site-diagonal) self-energy $\Sigma_n^{\sigma}$ as, 
\begin{equation}
\Sigma_n^{\sigma} \ll d_{i\sigma} , d_{j\sigma}^{\dagger} \gg_n = 
U \sum_{\eta} \ll d_{i\sigma} f_{i\eta}^{\dagger} f_{i\eta} , 
d_{j\sigma}^{\dagger} \gg_n
\label{eomGerd4}
\end{equation}
we obtain from (\ref{eomGerd2}), 
\begin{eqnarray}
\Sigma_n^{\sigma} \left( (\mu + i\omega_n) \ll d_{i\sigma} , d_{i
\sigma}^{\dagger} \gg_n + \sum_j t_{ij} \ll d_{i\sigma} , d_{j
\sigma}^{\dagger} \gg_n \right) = \nonumber \\
= U \sum_{\eta} \langle
f_{i\eta}^{\dagger} f_{i\eta} \rangle + U \Sigma_n^{\sigma} \ll
d_{i\sigma} , d_{i\sigma}^{\dagger} \gg_n
\label{eomGerd5}
\end{eqnarray} 
Denoting the total average f-occupation per site $i$ by 
$w_1 = \sum_{\eta} \langle f_{i\eta}^{\dagger} f_{i\eta} \rangle$, 
and using
\begin{equation}
(\mu + i\omega_n) \ll d_{i\sigma} , d_{i
\sigma}^{\dagger} \gg_n + \sum_j t_{ij} \ll d_{i\sigma} , d_{j
\sigma}^{\dagger} \gg_n = 1 + \Sigma_n^{\sigma} \ll d_{i\sigma} ,
d_{i\sigma}^{\dagger} \gg, 
\end{equation}
which follows from Eq.~(\ref{eomGerd1}), we find, 
\begin{equation}
\Sigma_n^{\sigma} \left(1 - (U - \Sigma_n^{\sigma}) \ll
d_{i\sigma},d_{i\sigma}^{\dagger} \gg_n \right) = U w_1 , 
\label{Sigmafunctional2}
\end{equation}
which is equivalent  to Eq.~(\ref{sigma-CPA}) 
and which we recognize as the standard Hubbard-III (CPA) 
self-consistency equation for the self-energy. 
The fact that the exact d-electron self energy is given by the CPA, 
which becomes exact in the limit of infinite dimensions for disordered 
systems, has a simple physical interpretation: 
As the f-electron number per site is conserved, the d-electrons ``see'' 
an effective disordered alloy potential because at a certain site $i$ 
there is either the on-site potential 0, if the site is not occupied 
by an f-electron, i.e. with probability $w_0$, 
or there is the on-site potential $U$, if the site is occupied 
by an f-electron, i.e. with probability $w_1 = 1 - w_0$. 
However, the self-energy functional depends explicitly on $w_1$, 
which is not known unless one calculates the full partition function. 

\section*{Formalism for the f-electron Green's function}

The atomic f-propagator, $F(\tau)$, is defined 
in the interaction representation for $\tau > 0$ as (Brandt and Urbanek 1992), 
\begin{equation}
F_\eta(\tau)
                      \label{first}
=
- {\rm Tr}_{df}
\left\{
e^{ -\beta H_{at}}S(\lambda)
f_\eta(\tau)f_\eta^\dag(0)
\right\}
/{{\cal Z}_{at}(\lambda)},
\end{equation}
where the trace is taken over the atomic d- and f-states,  
including the spin labels, and $H_{at}$ is the full atomic 
Hamiltonian defined by Eq.~(\ref{H_atom}).
The time evolution of the f- and d- operators is now  defined as, 
$ f_\eta(\tau) = e^{\tau H_{at} } f_\eta \; e^{-\tau H_{at} }$,
and 
$d_\sigma^\dag(\tau)=e^{\tau H_{at} } d_\sigma^\dag \; e^{-\tau H_{at} }$, 
which leads to the EOM, 
\begin{equation}
                                                    \label{d_d_tau}
\frac{d}{d \tau} d_\sigma^\dag(\tau)
= 
[-\mu + U \sum_\eta f_\eta^\dag(\tau) f_\eta(\tau)] d_\sigma^\dag(\tau),
\end{equation}
\begin{equation}
                                                    \label{d_f_tau}
\frac{d }{d \tau} f_\eta(\tau)
= 
[-(E_f-\mu) - U \sum_\sigma d_\sigma^\dag(\tau) d_\sigma(\tau)] f_\eta(\tau).
\end{equation}
The integral representation of (\ref{d_d_tau}) and (\ref{d_f_tau}) is, 
\begin{equation}
                                                    \label{d_tau}
d_\sigma^\dag(\tau) 
= 
e^{-\mu\tau} 
T_\tau
e^{U \sum_\eta
\int_0^{\tau } d \tau^{''} f_\eta^\dag(\tau^{''}) f_\eta(\tau^{''})
}d_\sigma^\dag ,
\end{equation}
\begin{equation}
                                                    \label{ff_tau}
f_\eta(\tau) 
= 
e^{-(E_f-\mu)\tau}
T_\tau
e^{-U \sum_\sigma 
\int_0^{\tau} d \tau^{''} 
d_\sigma^\dag(\tau^{''}) d_\sigma(\tau^{''})}
f_\eta,
\end{equation}
which can be written as, 
\begin{equation}
                                                    \label{fff_tau}
f_\eta(\tau) 
=
e^{-(E_f-\mu)\tau} 
S'(\tau) f_\eta.
\end{equation}
Here $S'(\tau)$ is the time-ordered exponential, 
\begin{equation}
                                                    \label{S'}
S'(\tau)  
= 
T_\tau
\exp \left\{ 
      - \sum_\sigma 
         \int_0^{\beta} \int_0^\beta d \tau^\prime d \tau^{\prime\prime} 
                \chi_\tau(\tau^\prime,\tau^{\prime\prime})
                  d_\sigma^\dag(\tau^\prime)d_\sigma(\tau^{\prime\prime})
     \right\}  ,
\end{equation}
and 
\begin{equation}
                                             \label{chi_tau}
\chi_\tau(\tau^\prime,\tau^{\prime\prime})
=
U \Theta(\tau-\tau{'}) \delta(\tau^{'} - \tau^{''}),
\end{equation} 
with $\Theta(x)$ the unit step function $\Theta(x>0)=1$ and $\Theta(x<0)=0$.
Eqs.~(\ref{d_tau}) and (\ref{fff_tau}) lead to the expression, 
\begin{equation}
F_\eta(\tau)
                                                    \label{SS'}
=-
e^{-(E_f-\mu)\tau} T_\tau
{\rm Tr}_{d\,f}
\left\{
e^{ -\beta H_{at}} 
S(\lambda)
S'(\chi_\tau)
f_\eta f_\eta^\dag
\right\}
/{{\cal Z}_{at}(\lambda)},
\end{equation}
which shows that the operator $f_\eta f_\eta^\dag(0)$ and the constraint 
$n_f=\sum_\eta f_\eta f_\eta^\dag\le 1$ eliminates
all of the $n_f=1$ states from the trace in (\ref{SS'}). 
In addition, all of the intermediate states of the system, 
obtained by applying $S\,S'$ to an initial state in the $n_f=0$ subspace,  
remain in the same subspace. This is because the f-operators 
in $S$ and $S'$ always appear as equal-time pairs, 
$f_\eta^\dag(\tau')f_\eta(\tau')$ and just count the f-electrons 
at time $\tau'$. Thus, the statistical problem for $F_\eta(\tau)$
is restricted to a constant $n_f$ subspace and we can replace the 
operator expression $\sum_\eta=f_\eta^{\dag}f_\eta$ in $H_{at}$ 
and in $S\,S'$ by its eigenvalue 0 or 1. 
We use $n_f=0$ for $\tau>0$ propagation and $n_f=1$ for $\tau<0$ propagation. 
The f-electron Green's function for $\tau\geq 0$ becomes, 
\begin{equation}
F_\eta(\tau)
                                                    \label{SxS'}
=-
e^{-(E_f-\mu)\tau}
{\rm Tr}_{d}
\left\{
e^{ -\beta H_{0}} 
S(\lambda)
S'(\tau)
\right\}
/{{\cal Z}_{at}(\lambda)}.
\end{equation}
with the statistical operator defined once again by  $H_0$ in 
Eq.~(\ref{ham_nonint})
and the statistical averaging performed with respect to all 
possible states of a d-electron perturbed by the $\lambda$-field
and by the additional time-dependent field  $\chi_\tau$

In the interaction representation defined by $H_0$, the time dependence 
of annihilation and creation operators is again 
$d_\sigma(\tau)=\exp (\mu\tau)d_\sigma$ and 
$d_\sigma^\dag(\tau)=\exp(-\mu\tau)d_\sigma^\dag$,
and the time-ordering becomes trivial. 
Thus, the product of two T-ordered exponentials in the expression 
(\ref{SxS'}), can be written as a single T-ordered exponential, 
\begin{equation}
S(\tilde\lambda_\sigma)
=
T_\tau
\exp
\left\{ -\sum_\sigma
\int_0^\beta\int_0^\beta \; d \tau^{'} d \tau^{''}
\tilde{\lambda_\sigma}(\tau^{'},\tau^{''})
d_\sigma^\dag(\tau^{'}) d_\sigma(\tau^{''})
\right\}.
\end{equation}
where 
\begin{equation}
{\tilde\lambda_\sigma}(\tau^{\prime},\tau^{\prime\prime})
=
\lambda^\sigma(\tau^\prime,\tau^{\prime\prime}) 
- 
\chi_\tau(\tau^\prime,\tau^{\prime\prime}), 
\end{equation}
which also depends on the external time $\tau$.
The f-electron Green's function becomes, 
\begin{equation}
                                        \label{final}
F_\eta(\tau)
= -
e^{-(E_f-\mu)\tau} 
T_\tau {\rm Tr}_d
\left\{
e^{-\beta H_0} 
S({\tilde\lambda})
\right\}
/{{\cal Z}_{at}(\lambda)},   
\end{equation}
and the problem is reduced to the evaluation of the statistical 
sum of a single atomic d-level coupled to a time-dependent 
${\tilde\lambda}$-field.
\subsubsection*{Effective partition function for the f-electron problem}
The effective partition function required for the f-propagator 
can be written in the factorized form,
\begin{equation}
                                        \label{Z_tilde-factorized}
{\cal Z}_0({\tilde\lambda})
= 
T_\tau {\rm Tr}_d
\left\{
e^{-\beta H_0} S({\tilde\lambda})
\right\}
= 
\prod_\sigma {\cal Z}_0^\sigma(\tilde\lambda_\sigma).
\end{equation}
The factorization (\ref{Z_tilde-factorized}) holds because 
the time evolution due to $H_0$  is such that the annihilation 
and creation operators with different spin-labels commute, 
regardless of their time arguments, 
and the exponential operator $S({\tilde\lambda})$ can be factorized. 

The time-translation-invariant component of the $\tilde\lambda$-field 
is determined by mapping the FK atom onto the FK lattice, 
while the additional $\chi_\tau$-component is defined by the Coulomb 
interaction between the f- and d-electrons during the time 
interval $(0,\tau)$. The presence of this 
additional time-dependent field can be understood as follows. 
In the FK atom the operator dynamics is defined by $H_{at}$ 
and the d-occupancy of the state vector is time-dependent 
because of the $\lambda$-field. 
The interaction between f- and d-electrons gives rise to 
fluctuations in the f-level position and makes the potential energy 
of the system time-dependent.
In the effective d-electron system described by Eq.~(\ref{final}),  
the changes in interaction energy due to the fluctuating f-d 
interaction energy are represented by the $\chi_{\tau}$-field. 
In other words, the f-electrons with an infinite lifetime,
acquire dynamics due to the coupling to the d-electron fluctuations.
In this respect, the FK problem is similar to the X-ray edge 
problem(Si et al., 1992). 
However, while the X-ray problem is formulated as a single site problem,
in the FK atom the self-consistently determined $\lambda$-field 
keeps track of all other f-sites in the lattice. At high temperatures,
where the coherent scattering of conduction electrons on the f-ions 
could be neglected, the single-site model (where there is no self-consistency
to determine the $\lambda$-field) might be representative of the 
lattice(Si et al., 1992). But at low temperatures, where 
coherence develops, the single-site description and the X-ray 
analogy might not be appropriate, and the lattice effects due 
to the $\lambda$--field have to be taken into account. 

The partition function ${\cal Z}_0({\tilde\lambda})$
cannot be calculated with the same procedure as for 
${\cal Z}_0(\lambda)$ because the  ${\tilde\lambda}$-field is no longer 
a function of $\tau-\tau^\prime$. Thus, the effective 
propagator associated with ${\cal Z}_0({\tilde\lambda})$ 
cannot be diagonalized by a Fourier transformation, and a simple 
differential equation for ${\cal Z}_0({\tilde\lambda})$ 
cannot be derived. 
Nonetheless, we use the functional differentiation of 
${\cal Z}_0({\tilde\lambda})$ with respect to ${\tilde\lambda}$
to generate an effective Green's function,  
\begin{equation}
                                          \label{g-functional_derivative}
g_\sigma(\tau^\prime,\tau^{\prime\prime}) 
=
 -\frac{ \delta \ln {\cal Z}^\sigma_0({\tilde\lambda}_\sigma) }
       {\delta {\tilde\lambda}_\sigma(\tau^{\prime\prime},\tau^\prime)}
\end{equation}
such that, 
\begin{equation}
                                            \label{g-definition}
g_\sigma(\tau^{\prime},\tau^{\prime\prime})
= -\frac{1}{ {\cal Z}^\sigma_0 ({\tilde\lambda_\sigma})}
{\rm Tr}_d \left<
T_{\tau}
e^{-\beta H_0^\sigma}
d_\sigma(\tau^{\prime}) d_\sigma^\dagger(\tau^{\prime\prime})
S({\tilde\lambda_\sigma})
\right>.
\end{equation}

Similarly, we introduce an auxiliary Green's function defined 
for a d-level driven by the $\chi_\tau$-field only, 
\begin{equation}
                                          \label{g_0-functional_derivative}
g_{0\sigma}(\tau^\prime,\tau^{\prime\prime}) 
=
 -\frac{ \delta \ln {\cal Z}^\sigma_0(\chi_\tau) }
       {\delta {\chi_\tau}(\tau^{\prime\prime},\tau^\prime)}
\end{equation}
where 
\begin{equation}
                                            \label{Z_0-chi-definition}
{\cal Z}^\sigma_0(\chi_\tau)
=
T_\tau {\rm Tr}_d
\left\{
e^{-\beta H_0} S(\chi_\tau)
\right\},
\end{equation} 
and  
\begin{equation}
                                            \label{g_0-definition}
g_{0\sigma}(\tau^\prime,\tau^{\prime\prime})
= -\frac{1}{ {\cal Z}^\sigma_0 ({\chi_\tau})}
{\rm Tr}_d \left<
T_{\tau}
e^{-\beta H_0^\sigma}
d_\sigma(\tau{^\prime}) d_\sigma^\dagger(\tau^{\prime\prime})
S({\chi_\tau})
\right>.
\end{equation}
The Green's functions $g_\sigma$ and $g_{0\sigma}$ depend 
explicitly on times $\tau^\prime$ and $\tau^{\prime\prime}$, 
and implicitly on $\tau$. 

The evaluation of ${\cal Z}^\sigma_0(\chi_\tau)$ and $g_{0\sigma}$ is 
straightforward because the evolution operator $S({\chi_\tau})$ 
does not change the number of d-electrons. The Hilbert space 
for d-states, in the absence of the $\lambda$-field, 
comprises only two states ($n_d=0$ and $n_d=1$), and  
the result for the partition function is simply, 
\begin{equation}
                                            \label{Z_0-chi-sigma}
{\cal Z}^\sigma_0(\chi_\tau)
=
1+e^{\beta\mu-U\tau}. 
\end{equation} 
and 
\begin{equation}
                                            \label{Z_0-chi}
{\cal Z}_0(\chi_\tau)
=
\prod_\sigma
{\cal Z}^\sigma_0(\chi_\tau).
\end{equation} 
The time-ordered product in Eq.~(\ref{g_0-definition}) has to be 
treated with some care, because the functional form 
of $\chi_\tau$-field is not the same in all the parts of 
the $(\tau^{'},\tau^{''})$-plane [see Eq.(\ref{chi_tau})].
Eventually, we obtain the following expressions 
for $g_{0\sigma}(\tau^{'},\tau^{''})$ (Brandt and Urbanek 1992), 
\begin{equation}
g_{0a^+}=
(\xi_0-1) e^{(\mu-U)\tau^{'}} e^{-(\mu-U)\tau^{''}},
\mbox{\hspace{3.5cm} for \hspace{1cm}}  
\tau^{''} < \tau^{'}< \tau,
\end{equation}         
\begin{equation}
g_{0a^-}=\xi_0     e^{(\mu-U)\tau^{'}} e^{-(\mu-U)\tau^{''}},
\mbox{\hspace{4.5cm} for \hspace{1cm}}  
\tau^{'} < \tau^{''}< \tau,
\end{equation}
\begin{equation}
g_{0b^+}= 
(\xi_0-1) e^{-U\tau} e^{\mu\tau^{'}} e^{-(\mu-U)\tau^{''}},
\mbox{\hspace{3.5cm} for \hspace{1cm}}  
\tau^{''}< \tau < \tau^{'},
\end{equation}
\begin{equation}
g_{0b^-}=
\xi_0     e^{U\tau} e^{(\mu-U)\tau^{'}} e^{-\mu\tau^{''}},
\mbox{\hspace{4.5cm} for \hspace{.5cm}}  
\tau^{'}< \tau < \tau^{''},
\end{equation}
\begin{equation}
g_{0c^+}=
(\xi_0-1) e^{\mu\tau^{'}} e^{-\mu\tau^{''}},
\mbox{\hspace{4.5cm} for \hspace{1cm}}  
\tau < \tau^{''}< \tau^{'},
\end{equation}
\begin{equation}
g_{0c^-}=
\xi_0     e^{\mu\tau^{'}} e^{-\mu\tau^{''}},
\mbox{\hspace{5.5cm} for \hspace{1cm}}  
\tau < \tau^{'}< \tau^{''} ,
\end{equation}
where the symbol $\pm$ in the subscript relates to evaluating
$g_{0}(\tau^{'},\tau^{''})$ 
above and below the line $\tau^{'}=\tau^{''}$ and the letter $a$, $b$, or $c$
refers to a different region on the $0-\beta$ square as depicted in 
Fig.~(\ref{integration_square}). Here the symbol
$\xi_0=1/(1+e^{U\tau-\beta\mu})$.  
The function $g_{0\sigma}(\tau^\prime,\tau^{\prime\prime})$ 
is implicitly time-dependent, because its functional form depends 
on the relative magnitude of $\tau$ with respect to 
$\tau^\prime$ and $\tau^{\prime\prime}$.
\begin{figure}[htb]
\center{\epsfig{file=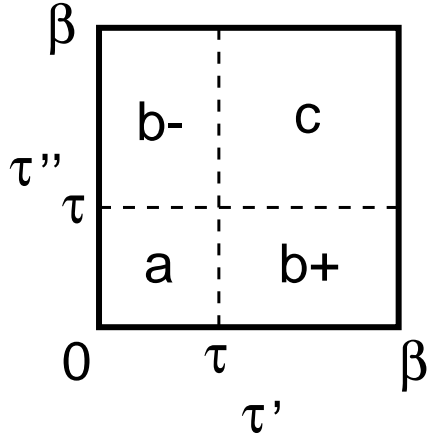,height=60mm}}
\caption{Different regions of the integration range for $\tau^\prime$ and
$\tau^{\prime\prime}$.}
                                      \label{integration_square}
\end{figure}

To proceed, we consider the periodic $\lambda$-field as an additional 
perturbation to the statistical problem defined by ${\cal Z}_{0}(\chi_\tau)$.
Thus, writing the full S-matrix in Eq.~(\ref{S'}) in the factorized 
form $S(\tilde\lambda)=S(\lambda) S(\chi_\tau)$, we find that 
the Green's functions obtained by the functional derivatives 
${ \delta \ln {\cal Z}({\tilde\lambda})}/
  {\delta {\tilde\lambda}(\tau^{\prime\prime},\tau^\prime)}$
and 
${ \delta \ln {\cal Z}_0(\chi_\tau) }/
       {\delta {\chi_\tau}(\tau^{\prime\prime},\tau^{\prime})}$
are related by a Dyson equation.
In the operator form, this can be written as, 
\begin{equation}
                                               \label{dyson_g}
g_\sigma
=
g_{0\sigma} -g_{0\sigma}\lambda^\sigma\,g_\sigma,
\end{equation} 
or, equivalently,  
\begin{equation}
                                               \label{dyson_g-1}
g_\sigma^{-1}
=
g^{-1}_{0\sigma} -\lambda^\sigma
=
g^{-1}_{0\sigma}(1-g_{0\sigma}\,\lambda^\sigma), 
\end{equation} 
where $g_\sigma$ and $g_{0\sigma}$ are non-diagonal integral operators 
both in the time and in the frequency representation, 
while $\lambda^\sigma$ is diagonal in the frequency representation.  
Next, we show that the partition function can be written as,
\begin{equation}
                                               \label{Z_0_det}
{\cal Z}^\sigma_0({\tilde\lambda^\sigma})
=
\det|g_\sigma^{-1}|. 
\end{equation}
This holds, because the functional derivatives
${\delta\ln{\cal Z}^\sigma_0({\tilde\lambda^\sigma})}
/{\delta{\tilde\lambda^\sigma}(\tau^\prime,\tau)}$
and 
${\delta\ln{\cal Z}^\sigma_0({\tilde\lambda^\sigma})}
/{\delta{\lambda^\sigma}(\tau^\prime,\tau)}$ 
define the same Green's function, $g(\tau,\tau^\prime)$, 
so that we can write, 
\begin{equation}    
                                               \label{delta-ln-Z_00}
\delta\,\ln {\cal Z}^\sigma_0(\tilde\lambda^\sigma)
=
-
\int_0^{\beta}d\tau\int_0^{\beta} d\tau^{\prime}
g_\sigma(\tau,\tau^\prime)\,
      \delta\lambda^\sigma(\tau^\prime,\tau)
=
-
\int_0^{\beta}d\tau\int_0^{\beta} d\tau^{\prime}
g_\sigma(\tau,\tau^\prime)\,
      \delta\tilde\lambda^\sigma(\tau^\prime,\tau)
\end{equation}
where $g_\sigma(\tau,\tau^\prime)$ is given by (\ref{g-definition}).
In other words, the variation of 
$\ln {\cal Z}^\sigma_0(\tilde\lambda)$ is not changed if 
$\lambda^\sigma(\tau^\prime,\tau)$ is shifted with respect to some 
arbitrary but fixed surface $[g_0]^{-1}(\tau^\prime,\tau)$. 
Using ~(\ref{dyson_g-1}) we write, 
$\delta\tilde\lambda^\sigma =\delta(\lambda^\sigma-[g_0]^{-1})
=-\delta g_\sigma^{-1}$,  
and obtain, 
\begin{equation}    
                                               \label{delta-ln-Z_0}
\delta\,\ln {\cal Z}^\sigma_0(\tilde\lambda)
=-
\int_0^{\beta}d\tau\int_0^{\beta} d\tau^{\prime}
g_\sigma(\tau,\tau^{\prime})\,
      \delta(\lambda^\sigma-g_{0\sigma}^{-1})(\tau^{\prime},\tau)
=
\int_0^{\beta}d\tau (g_\sigma\,\delta g_\sigma^{-1})(\tau,\tau), 
\end{equation}
where in the last equation we arrived at the diagonal matrix 
elements of $(g_\sigma\,\delta g_\sigma^{-1})$ by carrying out 
the matrix multiplication of $g_\sigma(\tau,\tau^\prime)$ 
and $\delta g_\sigma^{-1}(\tau^\prime,\tau)$.
Since $g_\sigma$ is the inverse of $g_\sigma^{-1}$, Eq.~(\ref{delta-ln-Z_0})
can be written as, 
\begin{equation}    
                                               \label{delta-ln-Z_0_g}
\delta\,\ln {\cal Z}^\sigma_0(\tilde\lambda)
=
\int_0^{\beta}d\tau
      \delta \ln [g_\sigma^{-1}](\tau,\tau),  
\end{equation}
which shows that $\delta\,\ln {\cal Z}^\sigma_0$ 
follows from the variation of Tr~$\ln [g_\sigma^{-1}]$. Thus, 
\begin{equation}    
                                               \label{ln-Z_0_g}
\ln {\cal Z}^\sigma_0(\tilde\lambda)
=
{\rm Tr}~\ln [g_\sigma^{-1}], 
\end{equation}
and the matrix identity ${\rm Tr}~ \ln A = \ln \det A$ leads 
to equation (\ref{Z_0_det}). 
The Dyson equation, Eq.~(\ref{dyson_g-1}), then yields the effective 
partition function as a continous determinant, 
\begin{equation}
                                               \label{det_g_1}
{\cal Z}^\sigma_0({\tilde\lambda})
=
\mbox{det} 
|g_{0\sigma}^{-1}|\;
\mbox{det} 
|1-g_{0\sigma}\,\lambda^\sigma|,
\end{equation}
which can be written as
\begin{equation}
                                               \label{det_g_Z}
{\cal Z}^\sigma_0({\tilde\lambda}_\sigma)
=
{\cal Z}^\sigma_0(\chi_\tau)
\;
\mbox{det} 
|1-g_{0\sigma}\,\lambda^\sigma|
=
{\cal Z}^\sigma_0(\chi_\tau)
\;
\mbox{det} D^\sigma. 
\end{equation}
Here, we introduced the integral operator $D^\sigma$, which is defined 
in the $\tau$--representation by its matrix elements 
\begin{equation}
                                               \label{D_tau_ij}
D^\sigma_{\tau_1,\tau_2}=\delta(\tau_1-\tau_2)
       - \int_0^\beta d\tau^{\prime\prime}
        g_{0\sigma}(\tau_1,\tau^{\prime\prime})
        \lambda^\sigma(\tau^{\prime\prime},\tau_2). 
\end{equation}
The Fourier transform of (\ref{D_tau_ij}) defines the integral operator 
$D^\sigma$ in frequency space where its matrix elements form
a discrete set. The time-translation invariance
of the $\lambda$-field leads to the expression, 
\begin{equation}
                                               \label{D_pq_ft}
D^\sigma_{pq}
=\delta_{pq} - 
\frac{\lambda^\sigma_q}{\beta}
\int_0^\beta d\tau^{'} e^{i\omega_p \tau^{'} }
\int_0^\beta d\tau^{''} g_{0\sigma}(\tau^{'},\tau^{''})
e^{-i\omega_q \tau^{''}}.
\end{equation}
where the integrals 
\begin{equation}
                                               \label{M_pq}
M^\sigma_{pq}
= 
\int_0^\beta d\tau^{'} e^{i\omega_p \tau^{'} }
\int_0^\beta d\tau^{''}e^{-i\omega_q \tau^{''}} 
g_{0\sigma}(\tau^{'},\tau^{''}) 
\end{equation}
are given in the Appendix. The implicit dependence of 
$g_{0\sigma}(\tau^{'},\tau^{''})$ on the external time
makes $M^\sigma_{pq}$ explicitly $\tau$--dependent.  
Since the determinant of an operator is the same in any basis, 
we can calculate ${\cal Z}^\sigma_0(\tilde\lambda)$ in the Matsubara 
representation using the discrete matrix elements (\ref{D_pq_ft}). 

The final form for the f-electron Green's function is thus (Brandt and Urbanek 1992), 
\begin{equation}      
                                               \label{F-final}
F_\eta(\tau)
=
-\frac{e^{-(E_f-U)\tau}\prod_\sigma [{\cal Z}_0^\sigma(\chi_\tau)\det 
D^\sigma_{pq}]}
{{\cal Z}_{at}(\lambda)}.
\end{equation}
It is useful to examine the expression in Eq.~(\ref{F-final}) for the
limits $\tau\rightarrow 0^+$ and $\tau\rightarrow \beta^-$ 
for the spinless case.  In the former
case we have $g_0(i\omega_p,i\omega_q)=\delta_{pq}/(i\omega_p+\mu)$ and
\begin{equation}
F(\tau=0^+)=-\frac{2e^{\beta\mu/2}\prod_n(i\omega_n+\mu-\lambda_n)/(i\omega_n)}
{{\cal Z}_{at}(\lambda)}=-w_0,
\end{equation}
and in the latter case we have $g_0(i\omega_p,i\omega_q)=\delta_{pq}
/(i\omega_p+\mu-U)$ and
\begin{equation}
F(\tau=\beta^-)=-\frac{e^{-\beta(E_f-\mu)}2e^{\beta(\mu-U)/2}\prod_n
(i\omega_n+\mu-\lambda_n-U)/(i\omega_n)}
{{\cal Z}_{at}(\lambda)}=-w_1,
\end{equation}
as we must have by directly evaluating the Green's function.
\section*{Formalism for the spontaneous hybridization}
Recent work by Sham and collaborators (Portengen et al 1996) proposed that the 
Falicov-Kimball model may have a ground state that has a nonzero 
average for a spontaneous hybridization $\langle df^{\dagger}\rangle$. 
Such a state would imply that the Falicov-Kimball fixed point is unstable
to the periodic Anderson model fixed point.  This instability is not
allowed at any finite temperature because the conservation of the local
f-electron number implies the existence of a local gauge symmetry which
cannot be broken at finite temperature due to Elitzur's 
theorem(Elitzur, 1975; Subrahmanyam and Barma, 1988).  This is the same conclusion arrived at 
from perturbation theory(Czycholl, 1999) and exact 
diagonalization(Farkasov\v sk\'y, 1999).  Here we will show how to test these
ideas exactly in the 
infinite-dimensional limit. (A mapping of the Falicov-Kimball
model with a Lorentzian density of states onto the X-ray edge problem 
shows that the hybridization susceptibility can diverge at 
$T=0$(Si et al., 1992), but it isn't clear that this behavior will
survive when one examines a conventional lattice.)

The static susceptibility for spontaneous hybridization satisfies
\begin{equation}
                                      \label{chi-hyb}
\chi_{hyb}=-\int_0^{\beta}d\tau
[{\rm Tr}_{df}\langle T_{\tau}e^{-\beta H_{at}}S(\lambda) f(\tau)d^{\dagger}(\tau)
d(0)f^{\dagger}(0)\rangle/Z_{at}(\lambda)+G(\tau)F(\tau)],
\label{chi_hyb_def}
\end{equation}
for the spinless Falicov-Kimball model.
The integrand can be determined by simply taking the functional derivative of 
the f-electron Green's function with respect to $\lambda(\tau,0)$:
\begin{equation}
\chi_{hyb}=-\int_0^{\beta}d\tau\left [
\frac{\delta F(\tau)}{\delta \lambda(\tau,0)}+G(\tau)F(\tau)\right ]
=\int_0^{\beta}d\tau \frac{e^{-(E_f-\mu)\tau}{\cal Z}_0(\chi_\tau)}
{{\cal Z}_{at}
(\lambda)}\frac{\delta [1-g_0\lambda]}{\delta \lambda(\tau,0)}.
\end{equation}
But using the identity $\det A=\exp [{\rm Tr}~ \ln A]$ tells us
\begin{equation}
\frac{\delta [1-g_0\lambda]}{\delta \lambda(\tau,0)}=
-\det[1-g_0\lambda]\{ (1-g_0\lambda)^{-1}g_0\}_{0,\tau}.
\end{equation}
Substituting this result into the integral then gives
\begin{equation}
\chi_{hyb}=\int_0^{\beta}d\tau F(\tau)\int_0^{\beta}d\tau^\prime 
[1-g_0\lambda]^{-1}_{0\tau^\prime}g_0(\tau^\prime,\tau),
\label{chi_hyb_tau}
\end{equation}
where it should be noted that the auxiliary Green's function $g_0$ is 
evaluated with the $\chi_\tau$ field and therefore must be recalculated for
each value of $\tau$ in the integrand (since $\chi_\tau$ varies with $\tau$).

Now we introduce a Fourier transform for the $\tau^\prime$
variable to rearrange this result into the following final form
\begin{equation}
\chi_{hyb}=\int_0^{\beta}d\tau F(\tau) T\sum_{mn}[1-g_0\lambda]^{-1}_{mn}
g_0(i\omega_n,\tau),
\label{chi_hyb_final}
\end{equation}
where the Green's function is Fourier transformed with respect to one
coordinate only
\begin{equation}
g_0(i\omega_n,\tau)=T\int_0^{\beta}d\tau^\prime e^{i\omega_n\tau^{\prime}}
g_0(\tau^\prime,\tau).
\label{g0_mixed}
\end{equation}
The calculation of $\chi_{hyb}$ requires little additional work to what is
needed to calculate $F(\tau)$.  At each value of $\tau$ we need only invert
the matrix $[1-g_0\lambda]$ and perform the relevant vector product with $g_0$
and the matrix summation.  

It is interesting to evaluate the susceptibility in the limit $U\rightarrow 0$.
Here we have $F(\tau)=-e^{-\tau(E_f-\mu)}/(1+e^{-\beta(E_f-\mu)})$, the
auxiliary Green function becomes time-translation invariant, so the matrix
is diagonal $[1-g_0\lambda]^{-1}_{mn}=(i\omega_m+\mu)/(i\omega_m+\mu-\lambda_m)
\delta_{mn}$, and the partial Fourier transformed Green function becomes
$g_0(i\omega_n,\tau)=e^{i\omega_n\tau}/(i\omega_n+\mu)$. The formula for
the susceptibility in Eq.~(\ref{chi_hyb_final}) can now be evaluated
directly by performing the summation over the Matsubara frequencies to
yield
\begin{equation}
\chi_{hyb}(U=0)=\int d\epsilon \rho(\epsilon)\frac{f(\epsilon-\mu)-f(E_f-\mu)}
{E_f-\epsilon},
\label{chy_hyb_u=0}
\end{equation}
where $f(x)=1/(1+e^{\beta x})$ is the Fermi function.  As $T\rightarrow 0$,
the susceptibility will diverge whenever the chemical potential is
equal to $E_f$ because the Fermi factors will limit the integration to 
the region $\epsilon\le E_f$, which will cause the integral to
diverge logarithmically.  A more careful analysis shows that the
susceptibility will behave like
\begin{equation}
\chi_{hyb}(U=0)\rightarrow -\frac{1}{2}\rho(E_f)\ln T + {\rm constant},
\label{chy_hyb_limit}
\end{equation}
as $T\rightarrow 0$.  Since we expect the susceptibility to be larger in
the interacting case, this analysis is suggestive that the spontaneous
hybridization will continue to diverge for nonzero $U$ as well.

\section*{Numerical solutions}
The numerical implementation of the above procedure is depicted schematically
in Fig.~\ref{algorithm_loop} and described as follows: 
We start with an initial guess for the self energy $\;\Sigma^{\sigma}\;$   
and calculate the local propagator in (\ref{eq: gloc}). 
Using (\ref{Dyson}) we calculate the bare atomic propagator 
${G_{0n}^{\sigma}}$ and find ${\cal Z}_0$ and  ${\cal Z}_{at}$. 
Next we obtain $w_0$, $w_1$ and find $G_{n}^{\sigma}$ from (\ref{G-atomic}). 
Using ${G_{0n}^{\sigma}}$ and $G_{n}^{\sigma}$, we compute the 
atomic self energy  and iterate the set of equations starting with
the new self energy until it converges 
to the fixed point. 
\begin{figure}[htb]
\center{\epsfig{file=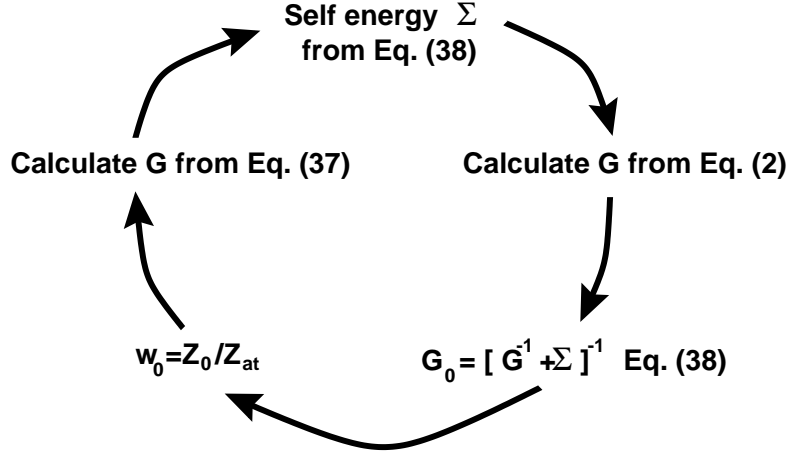,height=60mm}}
\caption{
Iterative algorithm for determining 
the self-consistent solution of
the local Green's function.}
                                      \label{algorithm_loop}
\end{figure}

The iterations on the  imaginary axis give static properties, like 
$n_f$,  and the static spin and charge susceptibilities. 
Having found the f-electron filling $w_1$
at each temperature, we iterate Eqs.~(\ref{eq: gloc}), (\ref{G-atomic}), 
and  (\ref{Dyson}) on the real axis and obtain the retarded dynamical 
properties, like the spectral function, the resistivity, 
the magnetoresistance, and the optical conductivity.  
At the fixed point, the spectral properties of the atom perturbed by 
$\lambda$-field coincide with the local spectral properties of the 
lattice.  

\subsubsection*{The f-electron spectrum and the results for 
                   classical intermediate-valence materials}

As an example, we consider first the d-electron and the f-electron 
Green's function for the spinless Falicov-Kimball model on a hypercubic
lattice, at half filling.
The interacting density of states for the conduction band is shown in 
Fig.~(\ref{zero_s_dos}), where $\rho_d(\omega)$ is plotted versus 
frequency for several values of $U$ in the high-temperature
homogeneous phase. The interacting density of states is independent 
of temperature at high temperatures but not at low temperatures 
where the system undergoes a phase transition to an AB ordered 
``chessboard'' phase~(Freericks and Lemanski, 2000). 
A metal-insulator  gap opens up at $U=1.5$.
\begin{figure}[htb]
\center{\epsfig{file=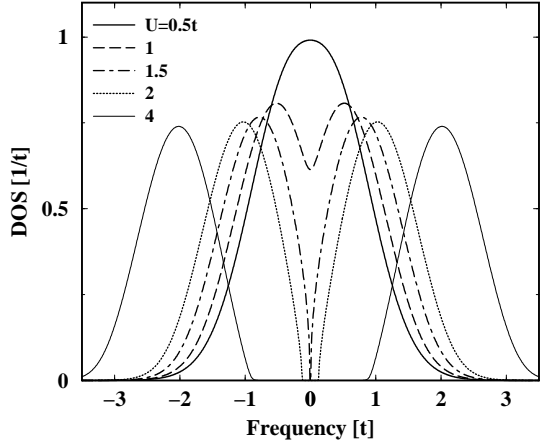,height=60mm,angle=0}}
\caption{
Interacting density of states plotted versus $\omega$ 
for several values of $U$, 
as indicated in the figure.}
                                      \label{zero_s_dos}
\end{figure}

We have illustrated how to calculate the f-electron Green's function,
and the result for the spinless Falicov-Kimball model on a hypercubic
lattice, at half filling is plotted in Fig.~(\ref{zero_S_f}) as a function 
of Matsubara frequency, and for several values of $U$. In the limit 
$U\rightarrow 0$, the Green's function becomes a noninteracting  delta-function
[which behaves like $1/(i\omega_n+\mu-E_f)$ on the imaginary axis].
\begin{figure}[htb]
\center{\epsfig{file=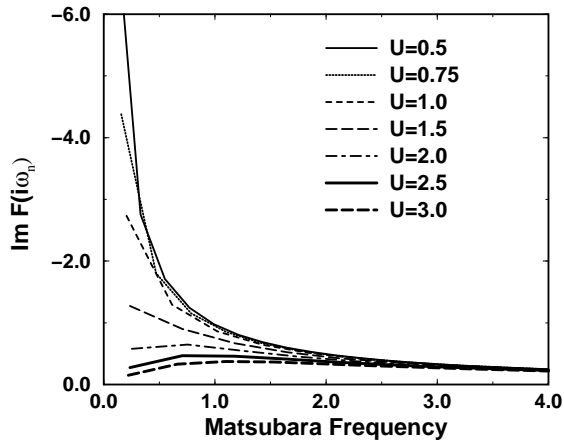,height=60mm}}
\caption{
F-electron Green's function is plotted versus Matsubara frequency, 
for several values of $U$ as indicated in the figure.}
                                      \label{zero_S_f}
\end{figure}
The sharp rise in the Green's function can be clearly seen for small $U$.
We expect that the Green's function will become a correlated insulator at
the same point that the conduction electron Green's function becomes insulating. 
Unfortunately,
the f-electron Green's function is temperature dependent here, and since we
are working at finite temperature we can only see the gap develop when $U$
becomes large [this is easiest to see by the fact that 
$F(i\omega_0)\rightarrow0$, which occurs for $U$ larger than about $2.0$].  

\begin{figure}[htb]
\center{\epsfig{file=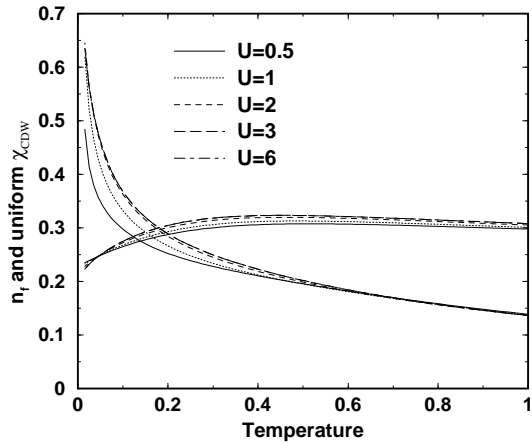,height=60mm}}
\caption{
The average f-electron concentration and
the uniform charge susceptibility are plotted versus temperature,
for several values of $U$ as indicated in the figure. Both quantities vary
little with the interaction strength.  The f-electron concentration is the
set of curves that decrease as $T\rightarrow 0$
and the susceptibility is the set of curves that
increase as $T\rightarrow 0$.}
                                                 \label{nf_cdw_n-0.5.eps}
\end{figure}

Recent calculations on the spin-one-half Falicov-Kimball model on
the Bethe lattice(Chung and  Freericks, 2000) have shown the existence of regions
of parameter space where the ground state is not a charge-density-wave
state or a phase separated state, but remains homogeneous down to $T=0$.  In
this region of the phase diagram, there are no other competing phases, so
the system is eligible to have a $T=0$ divergence of the spontaneous
hybridization.  Here we illustrate this behavior for the spinless model.
Previous calculations have found the possibility of spontaneous 
hybridization to be precluded by other phases(Farkasov\v sk\'y, 1999; Czycholl, 1999) 
or to occur for ``singular'' density of states(Si et al., 1992).  Here we
provide numerical evidence for the divergence at $T=0$ in a restricted 
region of parameter space for ``conventional lattices.''.
\begin{figure}[htb]
\center{\epsfig{file=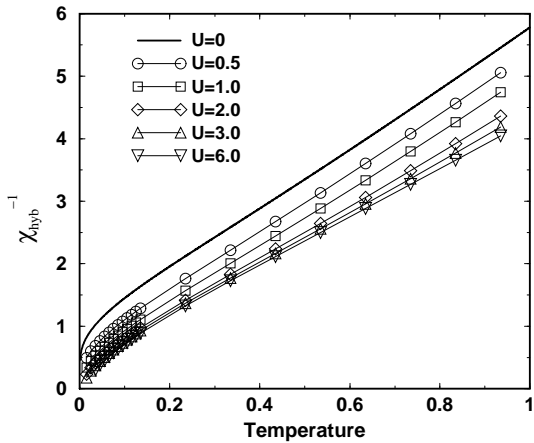,height=60mm}}
\caption{
The inverse of the hybridization susceptibility is
plotted versus temperature, for several values of $U$ 
as indicated in the figure.}
                                      \label{hyb_n-0.5.eps}
\end{figure}

We begin by finding a region of parameter space in the spinless FK
model where the classical intermediate
valence state is stable against phase separation
all the way down to $T=0$.  This is simplest
to find by repeating the previous analysis for the spin-one-half 
case(Chung and  Freericks, 2000) (we perform calculations on the Bethe lattice here).  
We can show that if we choose $n_{total}=0.5$ and $-1<E_f<0$,
then the intermediate-valence state is stable for small $U$.  We choose
$E_f=-0.75$, where the intermediate valence state appears to be stable
for all values of $U$ (we did not perform a Maxwell construction to check for
first-order phase transitions). 
A plot of the average f-electron concentration
and the uniform charge susceptibility versus temperature is given in 
Fig.~(\ref{nf_cdw_n-0.5.eps}).

Note how the f-electron concentration remains finite for all $U$ as $T
\rightarrow 0$ and how the compressibility remains finite as well.
The inverse of the hybridization susceptibility is shown
in Fig.~(\ref{hyb_n-0.5.eps}).
Note how the
logarithmic divergence at $U=0$ is difficult to see in this figure and 
how the inverse
susceptibility decreases as $U$ increases.  This then suggests that the
susceptibility will continue to diverge at $T=0$ for all $U$ (we know from
Elitzur's theorem that it cannot diverge at any finite $T$).

\subsubsection*{Results for YbInCu$_4$}

The numerical results for the spin-one-half FK model 
exhibit several features that one finds in the family of 
YbInCu$_4$ compounds. We consider here a hypercubic lattice with
a total electron filling of 1.5 and several values of $E_f$ and $U$,
such that $E_f>\mu (T=0)$, since that is the regime where the 
f-occupation can change rapidly as a function of $T$.

\begin{figure}[htb]
\center{\epsfig{file=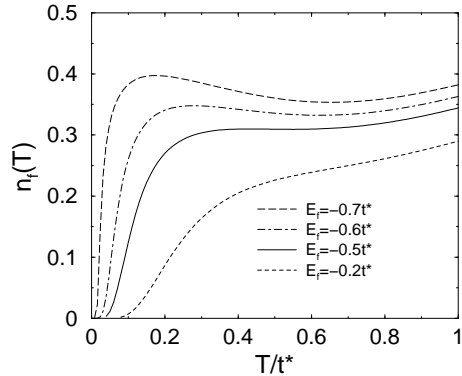,height=60mm,angle=-90}}
\caption{
Number of the f-holes plotted versus $T$
for $U=4$. The $E_f$ increases from
top to bottom, and is given by -0.7, -0.6, -0.5,
and -0.2, respectively.}
                                          \label{n_vs_T}
\end{figure}

The main results can be summarized in the following way. 
The occupancy of the f-holes at high temperatures is large and there is 
a huge magnetic degeneracy. The f-holes are energetically unfavorable 
but are maintained because of their large magnetic entropy. 
In Fig.~(\ref{n_vs_T}) we show $n_f$  
as a function of temperature, 
plotted for $U=4$, and $E_f$ ranging from -0.2 to -0.7. 
Below a certain temperature, which depends on $U$ and $E_f$,
there is a rapid transition from the high-temperature phase with a moderate
f-occupancy to the low-temperature phase where $n_f\rightarrow 0$. 
The ``transition'' occurs at a crossover temperature $T_v$ and
becomes sharper and is pushed to lower temperatures 
as $E_f$ decreases at fixed $U$. 
The uniform f-spin susceptibility is obtained by calculating the spin-spin 
correlation function(Brandt and Urbanek, 1992; Freericks and Zlati\'c, 1998) and is given by  
$\chi(T)=C n_f(T)/T$, where $C=g_L^2\mu_B^2 J(J+1)/3k_B$ is the Curie 
constant. Thus it is clear that in the high-temperature 
phase, where  $n_f(T)$ does not change much, the susceptibility 
approaches the Curie law. But as far as the f-d interaction give rise to changes 
in the f-occupancy the susceptibility assumes only the form of 
an approximate Curie-Weiss law. 
In the region where $n_f(T)$ changes rapidly, the susceptibility exhibits 
a sharp maximum, which separates the magnetic and non-magnetic 
regions of the phase diagram. 
Below $T_v$, the f-susceptibility is negligibly small, and the total 
susceptibility is due to conduction electrons and is Pauli like. 
The other static correlation functions have also been calculated, 
and the results obtained in the homogeneous phase are discussed 
in (Freericks and Zlati\'c 1998). 

The interacting density of states $\rho_d(\omega)$ for the conduction 
electrons, calculated for $U=4$ and $E_f=-0.5$, 
is plotted in Fig.~(\ref{dos_vs_omega}) versus frequency for several 
temperatures. (The zero of energy is measured with respect to $\mu$.) 
The  high-temperature density of states has a gap of the order of $U$, 
and the chemical potential is located within the gap. 
Below the crossover temperature $T_v$, $n_f$ is small, the correlation effects 
are reduced, and $\rho_d(\omega)$ assumes a nearly non-interacting 
shape, with large $\rho_d(\mu)$  
and halfwidth $W\simeq 1$.
\begin{figure}[htb]
\center{\epsfig{file=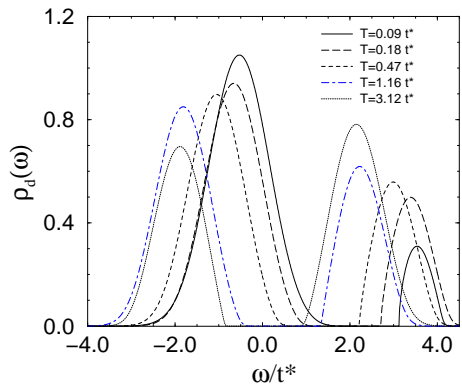,height=60mm,angle=-90}}
\caption{
Interacting density of states plotted versus 
$\omega$ for $U=4$, $E_f=-0.5$ 
($T_v=0.14$), and for various temperatures, 
as indicated in the figure.}
                                      \label{dos_vs_omega}
\end{figure}

The intraband optical conductivity is determined by an integral of the
spectral function~(Pruschke, et al., 1995) as
\begin{equation}
\sigma(\omega)=\sigma_0\int d\epsilon\rho(\epsilon)\int d\nu 
\frac{f(\nu)-f(\nu+\omega)}{\omega} A(\epsilon,\nu)A(\epsilon,\nu+\omega),
\end{equation}
where $A(\epsilon,\nu)=-\frac{1}{\pi}{\rm Im}~ [1/(\nu+\mu-\Sigma(\nu)-\epsilon)]$
is the spectral function. 
The result for $\sigma(\omega)$ obtained in such a way is plotted in 
Fig.~(\ref{optical_vs_omega}) 
as a function of frequency, for several temperatures. 
Above $T_v$, we observe a reduced Drude peak around $\omega=0$ 
and a pronounced high-frequency peak around $\omega\simeq U$. 
The shape of $\sigma(\omega)$ changes completely across $T_v$. 
Below $T_v$ the  Drude peak is fully developed and there 
is no high-energy (intraband) structure. 
However, if the renormalized f-level is close to $\mu$, 
the interband d-f transition could lead to an additional 
mid-infrared peak.
\begin{figure}[htb]
\center{\epsfig{file=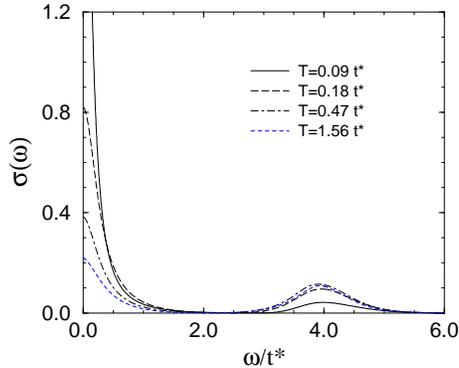,height=60mm,angle=-90}}
\caption{Optical conductivity 
plotted versus $\omega/t^*$ 
for various temperatures.
The $U$, $E_f$, and $T_v$, are the same as 
in Fig.~(\ref{dos_vs_omega}).}
                                 \label{optical_vs_omega}
\end{figure}

If we estimate the f-d correlation in YbInCu$_4$ from the 8000 $cm^{-1}$ 
peak in the optical conductivity data(Garner et al., 2000),  
we obtain the experimental value $U\simeq 1$ eV. Together with 
$T_v=42$ K (Sarrao et al 1999) this gives the ratio $U/T_v\simeq 200$. 
If we take $U=4$ and adjust $E_f$ so as to bring 
the theoretical value of $T_v$ in 
agreement with the thermodynamic and transport data on YbInCu$_4$, 
we get a high-frequency peak in $\sigma(\omega)$ at about 
8000 $cm^{-1}$, 6000 $cm^{-1}$, and 1500 $cm^{-1}$,   
for $E_f=-0.75$, $E_f=-0.7$, and $E_f=-0.5$, respectively.

From the preceding discussion it is clear that the Falicov-Kimball model 
captures the main features of the experimental data for YbInCu$_4$ 
and similar compounds. 
However, our calculations describe much better the doped Yb systems with broad 
transitions, than those compounds which show a first-order transition. 
The numerical curves can be made sharper
(by adjusting the parameters) but they only become discontinuous 
in a narrow parameter range. 
Our results indicate that the temperature- and field-induced anomalies 
are related to a metal-insulator transition, 
which is caused by a large FK interaction and triggered by 
the temperature- or the field-induced change in the f-occupancy. 

\section*{Summary and outlook}
We studied the static and dynamic correlation functions of the 
infinite--dimensional FK model by an equation-of-motion method. 
The exact solution (Brandt and Mielsch, 1989; Brandt and Urbanek, 1992) 
has been presented for the model with an arbitrary number of electrons, and for a 
$(2s+1)$--fold degenerate d-state and a $(2S+1)$--fold degenerate d-state. 
In the large-U limit, and for a range of parameters, the spin-one-half 
model exhibits a transition from a high-temperature semiconductor or a 
semimetal, with well defined f-moments, to a low-temperature Pauli metal. 
The static and dynamic correlation functions show many similarities 
with the experimental data on valence fluctuating Yb-compounds, 
and perhaps on SmB$_6$, but the crossovers calculated for the 
spin-one-half model are less sharp than what is seen in the experimental data. 
We believe, the sharpness of the transition in Yb--compounds is due to the large 
entropy change, when Yb ions switch from the magnetic $4f^{13}$ configuration 
(with a 14--fold degenerate f-hole in the $J=7/2$ spin-orbit state) 
to a non-magnetic $4f^{14}$ configuration. 
The model with a 14-fold degenerate f-state and a 2-fold degenerate d-state 
can easily be solved by the methods explained in this paper, and we expect 
to see a much sharper transition there. The numerical analysis of such a model, 
and the study of the correlation functions for various values of the 
ratio $(2S+1)/(2s+1)$, will be the subject of subsequent work.

A more serious difficulty with the FK model is that it neglects the quantum 
fluctuations of the f-state and considers only statistical fluctuations. 
That is, the lifetime of an f-state is assumed to be infinite, and the 
width of the f-spectrum arises only because the f-electrons couple to density 
fluctuations in the conduction band. Thus, the valence transition in the FK model 
is accompanied by a substantial change in the f-occupancy, and the loss of the 
magnetic moment is associated with the loss of f-holes. 
In real materials, the loss of the local moment seems to be due to quantum 
fluctuations and lifetime effects, rather than to the disappearance of the f-holes, 
The description of the quantum valence fluctuating ground state would have to 
consider the hybridization between the f- and d-states, and that would require 
a periodic Anderson model with an additional FK interaction. 
The EOM method elaborated in this paper does not produce the solution for such a 
generalized model. 

The actual situation pertaining to Yb ions in the mixed-valence state might 
be quite complicated,  since one must consider 
an extremely asymmetric limit of the Anderson model, 
in which the ground state is not Kondo-like, there is no Kondo resonance, 
and there is no single universal energy scale which is relevant at all 
temperatures(Krishnamurti et al., 1980).
We speculate that the periodic Anderson model 
with a large FK term will exhibit the same behavior 
as the FK model at high temperatures. Indeed, if the conduction band and the 
f-level are gapped, and the width of the f-level is large, then 
the effect of the 
hybridization can be accounted for by renormalizing the parameters
of the FK model. On the other hand, if the low-temperature 
state of the full model is close to the valence-fluctuating fixed point 
with the conduction band and hybridized f-level close to the Fermi 
level, then the likely effect of the FK correlation is to renormalize the 
parameters of the Anderson model. 

This picture is borne out from our examination of the spontaneous
hybridization for the spinless FK model on the Bethe lattice.  We find that
as $T$ is lowered the system seems to have a logarithmic divergence in the
spontaneous hybridization susceptibility at $T=0$. Normally we cannot
reach such a phase because the system will have a phase transition
to either a phase separated state or a charge-density wave, but we can tune
the system so that it remains in a classical intermediate valent state down
to $T=0$.  When this occurs, effects of even a small hybridization
will take the system away from the FK fixed point at low enough $T$
and hybridization can no longer be neglected.

\section*{Acknowledgments}
This research was supported by the National Science Foundation under grant DMR-9973225. 
V. Z. acknowledges the hospitality of the Physics Department, University 
of Bremen, where a part of this work was completed. 
\section*{Appendix}
The matrix elements in Eqs.~(\ref{D_pq_ft}) and (\ref{M_pq}), which are used to 
calculate the determinant in Eq.~(\ref{F-final}) for the f-propagator, 
are given by the expressions,
\begin{eqnarray}
M_{(n\neq m)} &=& \xi_0 A_{nm} + (\xi _0-1) B_{nm} \\
M_{(n,n)}     &=& \xi_0 C_{nn} + (\xi _0-1) D_{nn}
\end{eqnarray}
where $\xi_0=1/(1+e^{U\tau-\beta\mu})$, and 
\begin{eqnarray}
A_{mn}
&=&  \frac{1}{ i(\omega_m-\omega _n)(i\omega_m+\mu -U) }
    -
     \frac{ e^{i\tau \omega_n+\mu \tau -\beta \mu } }
          { (i\omega _m+\mu )(i\omega_n+\mu ) }         
                                                            \nonumber\\     
&+&  \frac{  (e^{-(i\omega_m+\mu )\tau }+e^{-\beta \mu})
            (e^{(i\omega_n+\mu )\tau }-e^{U\tau}) }
          { (i\omega_m+\mu )(i\omega _n+\mu -U) }        
                                                            \nonumber\\
&-&  \frac{1}{ i(\omega_m-\omega_n)(i\omega_n+\mu ) }
    +
     \frac{ e^{-i(\omega_m-\omega _n)\tau } }
          { i(\omega_m-\omega_n)(i\omega_m+\mu ) }       
                                                            \nonumber\\
&+&  \frac{ e^{-(i\omega_m+\mu)\tau}[i(\omega_m-\omega_n) 
            e^{U\tau }-(i\omega_m+\mu-U)
            e^{(i\omega_n+\mu)\tau }] }
          { i(\omega_m-\omega_n)(i\omega_m+\mu-U)(i\omega_n+\mu-U) }
\left. \right.  
                                                            \nonumber\\
\end{eqnarray}
\begin{eqnarray}
B_{mn}
&=&   \frac{-1}{ i(\omega_m-\omega_n)(i\omega_n+\mu-U) }
    -
     \frac{  e^{-i(\omega_m-\omega_n)\tau } }
         { i(\omega_m-\omega_n)(i\omega_n+\mu) }
                                                            \nonumber\\
&-& \frac{  e^{i\omega_n\tau}(e^{-i\omega _m\tau }-e^{(\mu-U)\tau })
            -e^{\beta \mu }(e^{-(i\omega _m+\mu )\tau }
            -e^{-U\tau }) }
        { (i\omega _m+\mu-U)(i\omega _n+\mu) }
                                                            \nonumber\\
&+& 
      \frac{  i\omega_n+\mu-i(\omega_m-\omega_n)
             e^{\beta\mu-(i\omega_m+\mu )\tau } }
           {  i(\omega_m-\omega_n)(i\omega_m+\mu )(i\omega_n+\mu )}
                                                           \nonumber\\
&+& 
      \frac{  e^{-i(\omega_m-\omega_n)\tau }[i\omega_n+\mu-U
             +i(\omega_m-\omega _n)
              e^{(i\omega _m+\mu -U)\tau }]}
           { i(\omega_m-\omega_n)(i\omega_m+\mu-U)(i\omega_n+\mu -U)} 
                                                            \nonumber\\
\end{eqnarray}

\begin{eqnarray}
C_{nn}
&=&      \frac{  (i\omega _n+\mu -U)\tau 
               - 1
               + e^{-(i\omega _n+\mu -U)\tau } }
             { (i\omega _n+\mu -U)^2 }                
                                                       \nonumber\\
&+&      \frac{ 1-e^{-(i\omega _n+\mu -U)\tau} 
                - e^{-\beta\mu }(e^{(i\omega_n+\mu)\tau}
                - e^{U\tau})}
             { (i\omega _n+\mu )(i\omega _n+\mu -U)}  
                                                       \nonumber\\
&+&      \frac{  (i\omega _n+\mu )(\beta -\tau )
                - 1
                - e^{-\beta \mu +(i\omega_n+\mu)\tau }}
              {  (i\omega _n+\mu )^2}
                                                       \nonumber\\
\end{eqnarray}

\begin{eqnarray}
D_{nn}
&=&      \frac{ -(i\omega _n+\mu -U)\tau
               - 1
               + e^{(i\omega_n+\mu-U)\tau} }
             {(i\omega _n+\mu -U)^2}
                                                       \nonumber\\
&+&     \frac{ 1-e^{(i\omega_n+\mu-U)\tau}
               - e^{\beta\mu}
                (e^{-(i\omega_n+\mu )\tau }- e^{-U\tau })}
             { (i\omega _n+\mu )(i\omega _n+\mu -U) }
                                                       \nonumber\\
&+&     \frac{ (i\omega _n+\mu )(\tau -\beta )-1
                -e^{\beta\mu-(i\omega_n+\mu)\tau }}
             { (i\omega _n+\mu )^2 }
\end{eqnarray}

These expressions generalize the matrix elements obtained in (Brandt and Urbanek 1992) 
for the system with electron-hole symmetry.
We checked, that for $\mu=U/2$ our expressions agree with those of Brandt 
and Urbanek (1992) for $m\neq n$, but we find a slight inconsistency for 
the case $m=n$. The formulae given here, and those given by Brandt and Urbanek 
agree only if the minus sign in front of the $i\omega_m$ term which appears 
in the numerator of the first line of the expression for $M_{(m,m)}$ 
in Brandt and Urbanek (1992) is replaced by a plus sign. 


\begin{thebibliography}{99}

\bibitem{brandt.89}  
U. Brandt and C. Mielsch, 
Z. Phys. B {\bf 75}, 365 (1989);

\bibitem{brandt.92}  
U. Brandt and M. P. Urbanek, 
Z. Phys. B {\bf 89}, 297 (1992).

\bibitem{czycholl}
G. Czycholl, 
Phys. Rev. B {\bf 59}, 2642 (1999).

\bibitem{elitzur}
S. Elitzur, Phys. Rev. D {\bf 12}, 3978 (1975).

\bibitem{falicov.69}  
L. M. Falicov and J. C. Kimball, 
Phys. Rev. Lett. {\bf 22}, 997 (1969).
                   
\bibitem{farkasovsky}
P. Farkasov\v sk\'y, 
Z. Phys. B {\bf 104}, 553 (1997);
Phys. Rev. B {\bf 59}, 9707 (1999).

\bibitem{freericks.98}
J. Freericks and V. Zlati\'c,
Phys. Rev. B {\bf 58}, 322 (1998);
V. Zlati\'c and  J. Freericks,
in {\em Proc. NATO ARW Conference, Bled 2000},
edited by P. Prelovsek, S. Sarkar, J. Bonca
(North-Holland, Amsterdam, 2001).

\bibitem{romek.00}
J. K. Freericks and R. Lemanski, 
Phys. Rev. B 61, 13438 (2000). 

\bibitem{garner.00}                
S.R. Garner, J.N. Hancock, Y.W. Rodriguez, Z. Schlesinger,
B. Bucher, Z. Fisk, and J.L. Sarrao, Phys. Rev. B 62 (2000) R4778.

\bibitem{guntherodt.82} 
G Guntherodt, W. A. Thompson, F. Holtzberg and Z. Fisk,  
in {\em Valence Instabilities}, edited by P.Wachter and H. Boppart
(North-Holland, Amsterdam, 1982).

\bibitem{kadanoff-baym}
L.P. Kadanoff and G. Baym, {\em Quantum Statistical Physics}  
(W. A. Benjamin, Menlo Park, CA, 1962).

\bibitem{kww.80}  
H. B. Krishnamurti, J. W. Wilkins, and K. G. Wilson,
Phys. Rev. B {\bf 21}, 1044 (1980).

\bibitem{metzner_vollhardt} 
W. Metzner and D. Vollhardt, Phys. Rev. Lett. {\bf 62}, 324 (1989).

\bibitem{sham.96}
T. Portengen, Th. Oestreich and L.J. Sham, Phys. Rev. Lett. {\bf 76},
2284 (1996); Phys. Rev. {\bf 54}, 17452 (1996).

 \bibitem{jarrell_optical}
Th. Pruschke, M. Jarrell, and J. K. Freericks,
Adv. Phys.  {\bf 44} 187 (1995). 

\bibitem{sarrao.99} 
J.L. Sarrao, Physica B, {\bf 259\&261}, 129 (1999).

\bibitem{georges_fk}
Q. Si, G. Kotliar, and A. Georges,
Phys. Rev. B {\bf 46}, 1261 (1992).  

\bibitem{indians}
V. Subrahmanyam and M. Barma, J. Phys. C {\bf 21}, L19 (1988).  

\bibitem{wacher.85} 
P.Wachter and G. Travaglini, 
J. Mag. Mat. Mater. {\bf 47-48}, 423 (1985).

\bibitem{fk_ivm}
Woonki Chung and J. K. Freericks, 
Phys. Rev. Lett. {\bf 84}, 2461 (2000).

\end{thebibliography}
\end{document}